\documentclass[a4paper,journal,twocolumn,10pt,final]{IEEEtran}

\makeatother

\usepackage[none]{hyphenat}
\usepackage[utf8]{inputenc}
\usepackage[T1]{fontenc}
\usepackage{textcomp}
\usepackage{scalefnt}
\usepackage{amsmath,amsfonts,amssymb,amsbsy,amstext}

\usepackage{mathtools}          
\usepackage{cmap}               
\usepackage{varwidth}

\let\originalleft\left
\let\originalright\right
\renewcommand{\left}{\mathopen{}\mathclose\bgroup\originalleft}
\renewcommand{\right}{\aftergroup\egroup\originalright}

\usepackage{graphicx}
\graphicspath{{.}{.\figs}}
\usepackage{placeins} 
\usepackage{float}
\usepackage{multirow}
\usepackage{tabularx}
\usepackage{booktabs}
\newcolumntype{C}{>{\centering\arraybackslash}X}
\newcolumntype{R}{>{\flushright\arraybackslash}X}
\newcolumntype{L}{>{\flushleft\arraybackslash}X}
\newcolumntype{P}{>{\centering\arraybackslash} p{0.5\linewidth}}

\usepackage[%
style=ieee,
backend=biber, 
url=false, 
]{biblatex}

\defbibheading{bibliography}[\refname]{\section*{#1}}

\makeatletter
\g@addto@macro{\UrlBreaks}{\UrlOrds}
\makeatother

\usepackage{acro}[=v2]

\NewDocumentCommand{\acro}{m o m o}
{%
	\IfValueTF{#2}{%
		\IfValueTF{#4}{%
			\DeclareAcronym{#1}{short={#2},long={#3},#4}
		}{%
			\DeclareAcronym{#1}{short={#2},long={#3}}
		}
	}{%
		\IfValueTF{#4}{%
			\DeclareAcronym{#1}{short={#1},long={#3},#4}
		}{%
			\DeclareAcronym{#1}{short={#1},long={#3}}
		}
	}
}

\acro{2D}{two-dimensional}
\acro{3D}{three-dimensional}
\acro{4D}{four-dimensional}
\acro{6D}{six-dimensional}
\acro{1G}{first generation}
\acro{2G}{second generation}
\acro{3G}{third generation}
\acro{4G}{fourth generation}
\acro{5G}{fifth generation}
\acro{5GC}{5G core network}
\acro{5G-StoRM}{5G stochastic radio channel for dual mobility}
\acro{6G}{sixth generation}
\acro{3GPP}{3rd generation partnership project}
\acro{3GPP2}{3rd generation partnership project 2}
\acro{A2A}{air-to-air}
\acro{A2G}{air-to-ground}
\acro{AA}{antenna array}
\acro{AC}{admission control}
\acro{AD}{attack-decay}
\acro{AE}{antenna element}
\acro{AF}{amplify and forward}
\acro{ABS}{almost blank subframe}
\acro{ACF}{autocorrelation function}
\acro{ADSL}{asymmetric digital subscriber line}
\acro{AHW}{alternate hop-and-wait}
\acro{AI}{artificial intelligence}
\acro{AMC}{adaptive modulation and coding}
\acro{AP}{access point}
\acro{APA}{adaptive power allocation}
\acro{API}{application protocol interface}
\acro{ARQ}{automatic repeat request}
\acro{AT}{averaged time}
\acro{ARMA}{autoregressive moving average}
\acro{ASE}{average squared error}
\acro{ASC}{adaptive satisfaction control}
\acro{ATES}{adaptive throughput-based efficiency-satisfaction trade-off}
\acro{AWGN}{additive white gaussian noise}
\acro{BAP}{backhaul adaptation protocol}
\acro{BB}{branch and bound}
\acro{BC}{branch and cut}
\acro{BD}{block diagonalization}
\acro{BER}{bit error rate}
\acro{BF}{best fit}
\acro{BL}{bit loading}
\acro{BLER}{block error rate}
\acro{BLPC-1}{bit loading with power constraint in hop 1}
\acro{BLPC-2}{bit loading with power constraint in hop 2}
\acro{BPC}{binary power control}
\acro{BPSK}{binary phase-shift keying}
\acro{BRA}{balanced random allocation}
\acro{BS}{base station}
\acro{BSP}{\acs*{BS} placement}
\acro{BSR}{buffer status report}
\acro{BoI}{band of interest}
\acro{C-link}{control link}
\acro{CAP}{combinatorial allocation problem}
\acro{CAPEX}{capital expenditure}
\acro{CB}{contextual bandit}
\acro{CBF}{coordinated beamforming}
\acro{CBR}{constant bit rate}
\acro{CBS}{class based scheduling}
\acro{CC}{congestion control}
\acro{CCL}{common cell list}
\acro{CDF}{cumulative distribution function}
\acro{CDL}{clustered delay line}
\acro{CDMA}{code-division multiple access}
\acro{CIR}{channel impulse response}
\acro{CH}{channel hardening}
\acro{CHO}{conditional handover}
\acro{C-RAN}{cloud-based radio access network}
\acro{CL}{closed loop}
\acro{CLT}{central limit theorem}
\acro{CLPC}{closed loop power control} 
\acro{CN}{core network}       
\acro{CNR}{channel-to-noise ratio}
\acro{CP}{control plane}
\acro{CPA}{cellular protection algorithm}
\acro{CPICH}{common pilot channel}
\acro{CoMP}{coordinated multi-point}
\acro{CQI}{channel quality indicator}
\acro{CRE}{cell range expansion}
\acro{CRM}{constrained rate maximization}
\acro{CRN}{cognitive radio network}
\acro{C-RNTI}{cell radio network temporary identifier}
\acro{CRRM}{centralized/common radio resource management}
\acro{CRS}{cell-specific reference signal}
\acro{CS}{coordinated scheduling}
\acro{CSI}{channel state information}
\acro{CSI-RS}{channel state information reference signal}
\acro{CTS}{clear to send}
\acro{CU}{centralized unit}
\acro{CUE}{cellular user equipment}
\acro{CWND}{congestion window size}
\acro{D2D}{device-to-device}
\acro{D2N}{drone-to-network}
\acro{DAA}{drone antenna array}
\acro{DAG}{directed acyclic graph}
\acro{DBS}{drone mounted base station}
\acro{DC}{dual connectivity}
\acro{DCA}{dynamic channel allocation}
\acro{DE}{differential evolution}
\acro{DF}{decode and forward}
\acro{DFT}{discrete fourier transform}
\acro{DIST}{distance}
\acro{DL}{downlink}
\acro{DMA}{double moving average}
\acro{DMRS}{demodulation reference signal}
\acro{D2DM}{\acs*{D2D} mode}
\acro{DMS}{\acs*{D2D} mode selection}
\acro{DP}{data plane}
\acro{DPC}{dirty paper coding}
\acro{DRA}{dynamic resource assignment}
\acro{DSA}{dynamic spectrum access}
\acro{DSM}{delay-based satisfaction maximization}
\acro{DU}{distributed unit}
\acro{E2E}{end-to-end}
\acro{ECC}{electronic communications committee}
\acro{EDF}{earliest deadline first}
\acro{EE}{energy efficiency}
\acro{EFLC}{error feedback based load control}
\acro{EI}{efficiency indicator}
\acro{e-ICIC}{enhanced inter-cell interference coordination}
\acro{eMBB}{enhanced mobile broadband}
\acro{EM}{electromagnetic}
\acro{eNB}{evolved node B}
\acro{EXP}{exponential}
\acro{EPA}{equal power allocation}
\acro{EPC}{evolved packet core}
\acro{EPS}{evolved packet system}
\acro{E-UTRAN}{evolved universal terrestrial radio access network}
\acro{ES}{exhaustive search}
\acro{FCP}{fundamental counting principle}
\acro{FCA}{flow control algorithm}
\acro{FD}{full duplex}
\acro{FDD}{frequency division duplex}
\acro{FDM}{frequency division multiplexing}
\acro{FDMA}{frequency division multiple access}
\acro{FER}{frame erasure rate}
\acro{FIFO}{first in first out}
\acro{FoV}{field-of-view}
\acro{FF}{fast fading}
\acro{FRS}[FS]{fast-rat scheduling}
\acro{FS}{fast switching}
\acro{FSB}{fixed switched beamforming}
\acro{FST}{fixed \acs*{SNR} target}
\acro{FT}{fourier transform}
\acro{FTP}{file transfer protocol}
\acro{Fwd}{forwarding}
\acro{GA}{genetic algorithm}
\acro{GAP}{generalized assignment problem}
\acro{GAP-MQ}{generalized assignment problem with minimum quantities}
\acro{GATES}{generalized adaptive throughput-based efficiency-satisfaction trade-off}
\acro{GBR}{guaranteed bit rate}
\acro{GLR}{gain to leakage ratio}
\acro{gNB}{gNodeB}
\acro{GOS}{generated orthogonal sequence}
\acro{GP}{gaussian process}
\acro{GPL}{GNU general public license}
\acro{GPS}{global positioning system}
\acro{GRP}{grouping}
\acro{GSM}{global system for mobile communications}
\acro{GTEL}{wireless telecommunications research group}
\acro{HARQ}{hybrid automatic repeat request}
\acro{HBF}{hybrid beamforming}
\acro{HCPP}{hardcore point process}
\acro{HetNet}{heterogeneous network}
\acro{HD}{half duplex}
\acro{HF}{high-frequency}
\acro{HH}{hughes-hartogs}
\acro{HardH}[HH]{hard handover}
\acro{HMS}{harmonic mode selection}
\acro{HO}{handover}
\acro{HOL}{head of line}
\acro{HPBW}{half power beamwidth}
\acro{HSDPA}{high speed downlink packet access}
\acro{HSR}{high-speed railway}
\acro{HSPA}{high speed packet access}
\acro{HTTP}{hypertext transfer protocol}
\acro{IAB}{integrated access and backhaul}
\acro{ICMP}{internet control message protocol} 
\acro{ICI}{intercell interference}
\acro{ICIC}{inter-cell interference coordination}
\acro{ID}{identification}
\acro{i.i.d.}{independent and identically distributed}
\acro{IETF}{internet engineering task force}
\acro{IPC}{individual power constraint}
\acro{UID}{unique identification}
\acro{IID}{independent and identically distributed}
\acro{IIR}{infinite impulse response}
\acro{ILP}{integer linear problem}
\acro{IMT}{international mobile telecommunications}
\acro{INV}{inverted norm-based grouping} 
\acro{IoT}{internet of things}
\acro{IP}{internet protocol}
\acro{IPv6}{internet protocol version 6}
\acro{IRA}{integrated resource allocation}
\acro{IRS}{intelligent reflecting surface}
\acro{ISD}{inter-site distance}
\acro{ISI}{inter symbol interference}
\acro{ISM}{industrial, scientific and medical}
\acro{ITU}{international telecommunication union}
\acro{JOAS}{joint opportunistic assignment and scheduling}
\acro{JOS}{joint opportunistic scheduling}
\acro{JP}{joint processing}
\acro{JRAPAP}{joint rb assignment and power allocation problem}
\acro{JS}{jump-stay}
\acro{JSM}{joint satisfaction maximization}
\acro{KKT}{karush-kuhn-tucker}
\acro{KPI}{key performance indicator}
\acro{LAC}{link admission control}
\acro{LA}{link adaptation}
\acro{LBS}{location based service}
\acro{LC}{load control}
\acro{LOS}{line of sight}
\acro{LP}{linear programming}
\acro{LTE}{long term evolution}
\acro{LTE-A}{\acs*{LTE}-advanced}
\acro{LTE-Advanced}{\ac{LTE-A}}
\acro{LTE-R}{\ac{LTE} railway}
\acro{LSP}{large scale parameter}
\acro{MADRDPG}{multi-agent deep recurrent deterministic policy gradient}
\acro{MeNB}{master \acs*{ENB}}
\acro{M2M}{machine-to-machine}
\acro{MAC}{medium access control}
\acro{MANET}{mobile ad hoc network}
\acro{MEDS}{method of exact doppler spread}
\acro{MC}{modular clock}
\acro{MCP}{minimal cost power}
\acro{MCS}{modulation and coding scheme}
\acro{MDB}{measured delay based}
\acro{MDI}{minimum \acs*{D2D} interference}
\acro{MDSM}{modified delay-based satisfaction maximization}
\acro{MDU}{max-delay-utility}
\acro{METIS}{mobile and wireless communications enablers for the twenty-twenty information society \acs*{5G}}
\acro{MF}{matched filter}
\acro{MFAP}{mobile femtocell access point}
\acro{MG}{maximum gain}
\acro{MH}{multi-hop}
\acro{mIAB}{mobile \ac{IAB}}
\acro{MILP}{mixed integer linear programming}
\acro{MIMO}{multiple input multiple output}
\acro{MINLP}{mixed integer nonlinear programming}
\acro{MIP}{mixed integer programming}
\acro{MISO}{multiple input single output}
\acro{MIT}{mobility interruption time}
\acro{ML}{machine learning}
\acro{MLWDF}{modified largest weighted delay first}
\acro{MME}{mobility management entity}
\acro{MMF}{max-min fairness}
\acro{mmMAGIC}{millimetre-wave based mobile radio access network for fifth generation integrated communications}
\acro{MMRP}{maximizing the minimal residual power}
\acro{MMRP-LB}{maximizing the minimal residual power with lower bound}
\acro{MMSE}{minimum mean square error}
\acro{mMTC}{massive machine-type communications}
\acro{mmWave}{millimeter wave}
\acro{MN}{master node}
\acro{MOS}{mean opinion score}
\acro{MPF}{multicarrier proportional fair}
\acro{MPRP}{maximization of the product of the residual powers}
\acro{MRA}{maximum rate allocation}
\acro{MR}{maximum rate}
\acro{MRC}{maximum ratio combining}
\acro{MRN}{mobile relay node}
\acro{MRT}{maximum ratio transmission}
\acro{MRUS}{maximum rate with user satisfaction}
\acro{MS}{mode selection}
\acro{MSE}{mean squared error}
\acro{MCG}{master cell group} 
\acro{MSI}{multi-stream interference}
\acro{MT}{mobile termination}
\acro{MTC}{machine-type communication}
\acro{IMS}{\acs*{IP} multimedia subsystem}
\acro{MTSI}{multimedia telephony services over \acs*{IMS}}
\acro{MTSM}{modified throughput-based satisfaction maximization}
\acro{MU-MIMO}{multi-user multiple input multiple output}
\acro{MU}{multi-user}
\acro{Multi-CUT}{multi-cell and multi-user and multi-tier}
\acro{NAS}{non-access stratum}
\acro{NB}{node B}
\acro{nBS}{neighbor base station}
\acro{NCL}{neighbor cell list}
\acro{NCR}{network-controlled repeater}
\acro{NG}{next generation}
\acro{NGC}{next generation core network}
\acro{NLP}{nonlinear programming}
\acro{NLOS}{non-line of sight}
\acro{NMSE}{normalized mean square error}
\acro{NN}{neural network}
\acro{NOMA}{non-orthogonal multiple access}
\acro{NORM}{normalized projection-based grouping}
\acro{NP}{non-polynomial time}
\acro{NR}{new radio}
\acro{NRT}{non-real time}
\acro{NSA}{non-stand-alone}
\acro{NSPS}{national security and public safety services}
\acro{OBF}{opportunistic beamforming}
\acro{OFDMA}{orthogonal frequency division multiple access}
\acro{OFDM}{orthogonal frequency division multiplexing}
\acro{OFPC}{open loop with fractional path loss compensation}
\acro{O2I}{outdoor-to-indoor}
\acro{OL}{open loop}
\acro{OLPC}{open-loop power control}
\acro{OL-PC}{open-loop power control}
\acro{OM}[O\&M]{operational \& maintenance}
\acro{OMA}{orthogonal multiple access}
\acro{OPEX}{operational expenditure}
\acro{ORA}{orthogonal resource allocation}
\acro{ORB}{orthogonal random beamforming}
\acro{JO-PF}{joint opportunistic proportional fair}
\acro{OSI}{open systems interconnection}
\acro{PA}{power allocation}
\acro{PAIR}{\acs*{D2D} pair gain-based grouping}
\acro{PAPR}{peak-to-average power ratio}
\acro{P2P}{peer-to-peer}
\acro{PBCH}{physical broadcast channel}
\acro{PBS}{pico base station}        
\acro{PC}{power control}
\acro{PCI}{physical cell \acs*{ID}}
\acro{PDCP}{packet data convergence protocol}
\acro{PDF}{probability density function}
\acro{PDU}{protocol data unit}
\acro{PER}{packet error rate}
\acro{PF}{proportional fair}
\acro{PGRA}{probabilistic graph based resource allocation}
\acro{P-GW}{packet data network gateway}
\acro{PHY}{physical}
\acro{PL}{pathloss}
\acro{PLR}{packet loss ratio}
\acro{PRABE}{power and resource allocation based on quality of experience}
\acro{PRB}{physical resource block}
\acro{PROJ}{projection-based grouping}
\acro{PSD}{power spectral density}
\acro{PSDe}{positive semi-definite}
\acro{ProSe}{proximity services}
\acro{PS}{packet scheduling}
\acro{PSO}{particle swarm optimization}
\acro{PSS}{primary synchronization signal}
\acro{PTAS}{polynomial-time approximation scheme}
\acro{PTP}{point-to-point}
\acro{PZF}{projected zero-forcing}
\acro{QAM}{quadrature amplitude modulation}
\acro{QHMLWDF}{queue-HOL-MLWDF}
\acro{QoE}{quality of experience}
\acro{QoS}{quality of service}
\acro{QPSK}{quadri-phase shift keying}
\acro{QSM}{queue-based satisfaction maximization}
\acro{QuaDRiGa}{quasi deterministic radio channel generator}
\acro{RACH}{random access}
\acro{RAISES}{reallocation-based assignment for improved spectral efficiency and satisfaction}
\acro{RAN}{radio access network}
\acro{RA}{resource allocation}
\acro{RAP}{rb assignment problem}
\acro{RAT}{radio access technology}[long-plural-form={radio access technologies}]
\acro{RATE}{rate-based}
\acro{RAU}{remote antenna unit}
\acro{RB}{resource block}
\acro{RBG}{resource block group}
\acro{REF}{reference grouping}
\acro{RET}{remote electrical tilt}
\acro{RF}{radio frequency}
\acro{RL}{reinforcement learning}
\acro{RLC}{radio link control}
\acro{RLF}{radio link failure}
\acro{RM}{rate maximization}
\acro{RMa}{rural macro}
\acro{RMJ-SNR}{region of the minimum joint snr}
\acro{RMEC}{rate maximization under experience constraints}
\acro{RMSE}{root-mean-square error}
\acro{RNC}{radio network controller}
\acro{RND}{random grouping}
\acro{RNN}{recurrent neural network}
\acro{RRA}{radio resource allocation}
\acro{RRM}{radio resource management}
\acro{RSCP}{received signal code power}
\acro{RSRP}{reference signal received power}
\acro{RSRQ}{reference signal received quality}
\acro{RR}{round robin}
\acro{RIS}{reconfigurable intelligent surfaces}
\acro{RRC}{radio resource control}
\acro{RSSI}{received signal strength indicator}
\acro{RT}{real time}
\acro{RTS}{request to send}
\acro{RU}{resource unit}
\acro{RUNE}{rudimentary network emulator}
\acro{RV}{random variable}
\acro{Rx}{receiver}
\acro{RZF}{regularized zero-forcing}
\acro{SA}{subcarrier assignment}
\acro{SAC}{session admission control}
\acro{sBS}{serving base station}
\acro{SC}{small cell}
\acro{SCI}{side control information}
\acro{SCon}[SC]{single connectivity}
\acro{SCG}{secondary cell group}
\acro{SCM}{spatial channel model}
\acro{SCS}{subcarrier spacing}
\acro{SC-FDMA}{single carrier - frequency division multiple access}
\acro{SCRV}{spatially consistent random variable}
\acro{SD}{soft dropping}
\acro{SDAP}{service	data adaptation protocol}
\acro{S-D}{source-destination}
\acro{SDPC}{soft dropping power control}
\acro{SDM}{space-division multiplexing}
\acro{SDMA}{space-division multiple access}
\acro{SE}{squared error}
\acro{SF}{shadow fading}
\acro{SMDP}{semi-markov	decision problem}
\acro{SeNB}{secondary \acs*{ENB}}
\acro{SER}{symbol error rate}
\acro{SES}{simple exponential smoothing}
\acro{S-GW}{serving gateway}
\acro{SIC}{successive interference cancellation}
\acro{SelfIC}[SIC]{self interference cancellation}
\acro{SINR}{signal to interference-plus-noise ratio}
\acro{SLNR}{signal to leakage-plus-noise ratio}
\acro{SNN}{strictly non-negative}
\acro{SI}{satisfaction indicator}
\acro{SelfI}[SI]{self interference}
\acro{SIP}{session initiation protocol}
\acro{SISO}{single input single output}
\acro{SIMO}{single input multiple output}
\acro{SIR}{signal to interference ratio}
\acro{SM}{subcarrier matching}
\acro{SMA}{simple moving average}
\acro{SMSE}{spatial-mean-square error}
\acro{SN}{secondary node}
\acro{SNR}{signal to noise ratio}
\acro{SON}{self organizing networks}
\acro{SoA}{state-of-the-art}
\acro{SoS}{sum-of-sinusoids}
\acro{SORA}{satisfaction oriented resource allocation}
\acro{SORA-NRT}{satisfaction-oriented resource allocation for non-real time services}
\acro{SORA-RT}{satisfaction-oriented resource allocation for real time services}
\acro{SP}{signal processing}
\acro{SPF}{single-carrier proportional fair}
\acro{SR}{smart repeater}
\acro{ASR}{advanced \ac{SR}}
\acro{SRA}{sequential removal algorithm}
\acro{SRB1}{signaling radio bearer~1}
\acro{SRS}{sounding reference signal}
\acro{SSB}{synchronization signal block}
\acro{SSP}{small scale parameter}
\acro{SSS}{secondary synchronization signal}
\acro{ST}{spanning tree}
\acro{STTD}{space time transmit diversity}[long-plural-form={space time transmit diversities}]
\acro{SU-MIMO}{single-user multiple input multiple output}
\acro{SU}{single-user}
\acro{tBS}{target base station}
\acro{SVD}{singular value decomposition}
\acro{TCP}{transmission control protocol}
\acro{TDD}{time division duplex}
\acro{TDM}{time division multiplexing}
\acro{TDMA}{time division multiple access}
\acro{TDMed}{time division multiplexed}
\acro{TETRA}{terrestrial trunked radio}
\acro{TP}{transmit power}
\acro{TPC}{total power constraint}
\acro{TR}{technical report}
\acro{TTI}{transmission time interval}
\acro{TTR}{time-to-rendezvous}
\acro{TTT}{time-to-trigger}
\acro{TSM}{throughput-based satisfaction maximization}
\acro{TU}{typical urban}
\acro{TV}{television}
\acro{TVWS}{\acs*{TV} white space}
\acro{Tx}{transmitter}
\acro{UAV}{unmanned aerial vehicle}
\acro{UABSC}{user-assisted bearer split control}
\acro{UDP}{user datagram protocol}
\acro{UDN}{ultra-dense networks}
\acro{UE}{user equipment}
\acro{UBA}{\ac{UE}-\ac{BS} association}
\acro{UEPS}{urgency and efficiency-based packet scheduling}
\acro{UFC}{federal university of cear\'{a}}
\acro{ULA}{uniform linear array}
\acro{UL}{uplink}
\acro{UMa}{urban macro}
\acro{UMi}{urban micro}
\acro{UMTS}{universal mobile telecommunications system}
\acro{UP}{user plane}
\acro{UPN}{user provided network}
\acro{URA}{uniform rectangular array}
\acro{URI}{uniform resource identifier}
\acro{URLLC}{ultra-reliable low-latency communications}
\acro{URM}{unconstrained rate maximization}
\acro{VBR}{variable bit rate}
\acro{VET}{variable electrical tilt}
\acro{VMR}{vehicle-mounted relay}
\acro{VR}{virtual resource}
\acro{VoIP}{voice over \acs*{IP}}
\acro{WCDMA}{wideband code division multiple access}
\acro{WF}{water-filling}
\acro{WID}{wireless infrastructure drone}
\acro{Wi-Fi}{wireless fidelity}
\acro{WiMAX}{worldwide interoperability for microwave access}
\acro{WINNER}{wireless world initiative new radio}
\acro{WLAN}{wireless local area network}
\acro{WMPF}{weighted multicarrier proportional fair}
\acro{WP}{work package}
\acro{WPF}{weighted proportional fair}
\acro{WSN}{wireless sensor network}
\acro{WWW}{world wide web}
\acro{WFQ}{weighted fair queuing}
\acro{XIXO}{(single or multiple) input (single or multiple) output}
\acro{XPR}{cross-polarization power ratio}
\acro{ZF}{zero-forcing}
\acro{ZMCSCG}{zero mean circularly symmetric complex gaussian}

\usepackage{datetime}
\usepackage{url}

\usepackage[caption=false,font=footnotesize]{subfig}

\usepackage{epstopdf}

\usepackage[hidelinks=true]{hyperref}

\usepackage[referable,flushleft]{threeparttablex}
\AtBeginEnvironment{tablenotes}{\footnotesize} 

\DeclareMathAlphabet{\mathppl}{T1}{ppl}{m}{it}
\DeclareMathAlphabet{\mathphv}{T1}{phv}{m}{it}
\DeclareMathAlphabet{\mathpzc}{T1}{pzc}{m}{it}

\newcommand{\SecRef}[2][]{Section#1~\ref{#2}}
\newcommand{\FigRef}[2][]{Fig.#1~\ref{#2}}

\newcommand{\TabRef}[2][]{Table#1~\ref{#2}}

\usepackage{standalone}
\usepackage{titlesec}
\titlespacing*{\section}
{0pt}{0.5\baselineskip plus 0.25\baselineskip minus 0.25\baselineskip}{0.25\baselineskip plus 0.125\baselineskip minus 0.125\baselineskip}
\titlespacing*{\subsection}
{0pt}{0.5\baselineskip plus 0.25\baselineskip minus 0.25\baselineskip}{0.25\baselineskip plus 0.125\baselineskip minus 0.125\baselineskip}

\AtBeginDocument{%
\abovedisplayskip 4pt plus 2pt minus 2pt
\belowdisplayskip \abovedisplayskip
\abovedisplayshortskip 2pt plus 1pt minus 1pt
\belowdisplayshortskip \abovedisplayskip
\textfloatsep 4pt plus 2pt minus 2pt
\intextsep 4pt plus 2pt minus 2pt
\floatsep 4pt plus 2pt minus 2pt
\dbltextfloatsep 4pt plus 2pt minus 2pt
\dblfloatsep 4pt plus 2pt minus 2pt
   
}

\begin{document}
\title{System Level Evaluation of Network-Controlled \\ Repeaters: Performance Improvement of Serving \\ Cell and Interference Impact on Neighbor Cells}

\author{Gabriel C. M. da Silva, Erik R. B. Falcão, Victor F. Monteiro, Darlan C. Moreira, \\ Diego A. Sousa, Tarcisio F. Maciel, Fco. Rafael M. Lima and Behrooz Makki. 
	\thanks{Behrooz Makki is with Ericsson Research, Sweden. The other authors are with the Wireless Telecommunications Research Group (GTEL), Federal University of Cear\'{a} (UFC), Fortaleza, Cear\'{a}, Brazil. Diego A. Sousa is also with Federal Institute of Education, Science, and Technology of Cear\'{a} (IFCE), Paracuru, Brazil. This work was supported by Ericsson Research, Sweden, and Ericsson Innovation Center, Brazil, under UFC.51 Technical Cooperation Contract Ericsson/UFC. The work of Victor F. Monteiro was supported by CNPq under Grant 308267/2022-2. The work of Tarcisio F. Maciel was supported by CNPq under Grant 312471/2021-1. The work of Francisco R. M. Lima was supported by FUNCAP (edital BPI) under Grant BP4-0172-00245.01.00/20.}%
}

\maketitle

\begin{abstract}
Heterogeneous networks have been studied as one of the enablers of network densification. %
These studies have been intensified to overcome some drawbacks related to propagation in \acp{mmWave}, such as severe path and penetration losses. %
One of the promising heterogeneous nodes is \ac{NCR}. %
It was proposed by the \ac{3GPP} in Release 18 as a candidate solution to enhance network coverage. %
In this context, this work performs a system level evaluation to analyze the performance improvement that an \ac{NCR} can cause in its serving cell as well as its interference impact on neighbor cells. %
Particularly, the results show a considerable improvement on the performance of \acp{UE} served by the \ac{NCR}, while neighbor \acp{UE} that receive the \ac{NCR} signal as interference are negatively impacted, but not enough to suffer from outage. %
\end{abstract}

\begin{IEEEkeywords}
	\acf{NCR}, wireless backhaul, coverage, \acf{5G}, \acf{6G}.
\end{IEEEkeywords}

\acresetall

\section{Introduction} \label{CHP:NCR_Interf_SEC:Intro}
Compared to the previous generation of wireless cellular systems, \ac{5G} networks have explored higher frequencies, e.g., \acp{mmWave}~\cite{Dong2022}. %
Some of the reasons for this interest are the fact that in \acp{mmWave} there are larger portions of available spectrum and that the required antenna arrays are smaller, which allows the deployment of more antenna elements creating narrow beams with high directional gain~\cite{Dong2022,Flamini2022,Polese2020}. %
Nonetheless, there are some disadvantages, e.g., suffering from high path and penetration losses~\cite{Flamini2022, Ayoubi2022}. %

One of the considered solutions to overcome the propagation losses is network densification. %
However, building from scratch a completely new wired infrastructure is expensive, takes time and, in some places, trenching may not be allowed, like historical areas. %
Then, nodes with wireless backhaul have emerged as a possible solution for the situations where wired backhaul is not a viable solution~\cite{Polese2020,Madapatha2020}. %

In this context, the present paper focuses on a new node with wireless backhaul called \ac{NCR}~\cite{Guo2022}. %
\ac{NCR} was introduced by \ac{3GPP} in Release 18~\cite{3gpp.38.867}. %
It is an enhanced version of traditional \ac{RF} repeaters. %
One of the novelties of \acp{NCR} is the fact that they have beamforming capability that can be controlled by a \ac{gNB} via side control information to improve the communication. %

In this paper, we introduce the concept of \ac{NCR} and study the performance of \ac{NCR}-assisted networks. %
More specifically, this paper performs a system level evaluation to analyze the performance improvement that an \ac{NCR} can cause in its serving cell. %
Also, we evaluate the effect of the interference that \acp{NCR} can have on the neighbor cells. %

The rest of the paper is organized as follows. %
\SecRef{CHP:NCR_Interf_SEC:NCR} presents the main concepts related to \ac{NCR}. %
The scenario considered in the performance evaluation is presented in \SecRef{CHP:NCR_Interf_SEC:Sys_Mod}. %
\SecRef{CHP:NCR_Interf_SEC:Perf_Eval} presents the performance evaluation results. %
Finally, conclusions are presented in \SecRef{CHP:NCR_Interf_SEC:Conclusion}. %

\section{Network Controlled Repeater} \label{CHP:NCR_Interf_SEC:NCR}
\begin{figure}[t]
	\centering
	\includegraphics[scale=0.35]{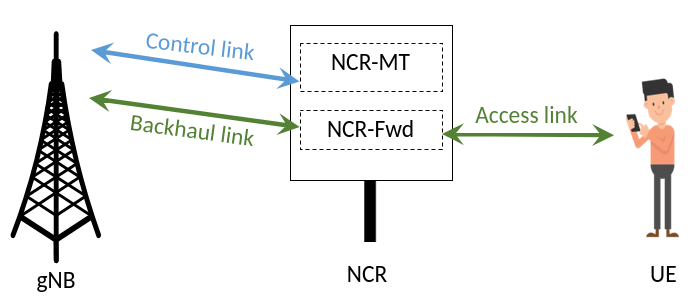}
	\caption{NCR split in NCR-MT and NCR-Fwd.}
	\label{FIG:NCR_Interf_SEC:NCR_MT_Fwd}
\end{figure}

Traditional \ac{RF} repeaters used in previous generations of wireless communications are non-regenerative nodes that amplify and forward a received signal. %
A direct consequence of this is the increase of interference in the system~\cite{qualcomm2021}. %

The \ac{NCR} concept is an evolution of the \ac{RF} repeaters that use side control information to overcome the negative aspects of traditional repeaters. %
Side control information that can be used by \acp{NCR} are~\cite{qualcomm2021}:
\begin{itemize}
	\item ON/OFF information: turn on/off the amplify and forward on a given slot;
	\item Timing information: dynamic \ac{DL}/\ac{UL} split;
	\item Spatial \ac{Tx}/ \ac{Rx}: beamforming capability.
\end{itemize}


\ac{NCR} can be split into \ac{NCR}-\ac{MT} and \ac{NCR}-\ac{Fwd}, as is shown in \FigRef{FIG:NCR_Interf_SEC:NCR_MT_Fwd}. %
The \ac{NCR}-\ac{MT} is responsible for exchanging side control information with its controlling \ac{gNB}. %
Its link is called control link, and it is based on \ac{NR} Uu interface. %
The \ac{NCR}-\ac{Fwd} is responsible for executing the \ac{AF} relaying~\cite{3gpp.38.867}. %



\section{System model} \label{CHP:NCR_Interf_SEC:Sys_Mod}

\begin{figure}[t]
	\centering
	\includegraphics[scale=0.5]{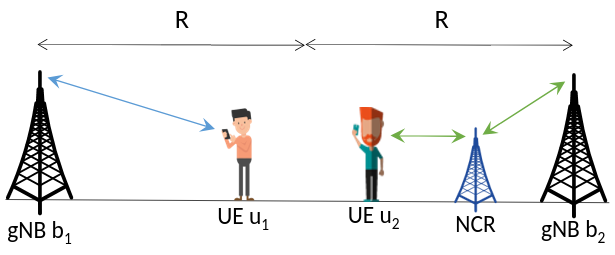}
	\caption{Scenario with two \acp{gNB} (distant 2$R$ from each other) and two \acp{UE}, connected to different \acp{gNB}. Moreover, it is considered that one of the \acp{gNB} controls an \ac{NCR}, which is deployed between its controller \ac{gNB} and its serving \ac{UE}.}
	\label{FIG:NCR_Interf_SEC:Scenario}
\end{figure}

As shown in \FigRef{FIG:NCR_Interf_SEC:Scenario}, consider a scenario with two \acp{gNB}, i.e., $b_{1}$ and $b_{2}$. 
The \ac{ISD} between them is equal to~$2R$. %
Also, consider that there is one \ac{UE} close to each cell edge, i.e., \acp{UE} $u_{1}$ and $u_{2}$ at cells of \acp{gNB} $b_{1}$ and $b_{2}$, respectively. %
In the cell of \ac{gNB} $b_{2}$, an \ac{NCR} is deployed between $b_{2}$ and \ac{UE} $u_{2}$ to enhance the link serving $u_{2}$ as shown in \FigRef{FIG:NCR_Interf_SEC:Scenario}. %

Since we consider that the \acp{UE} are close to the cell edge, the \ac{gNB} antenna array is deployed such that it steers to the middle point between its respective \ac{UE} and itself. %
Furthermore, the \ac{NCR} backhaul antenna array is always pointing towards the \ac{gNB} $b_{2}$ direction, while its access antenna array points towards the \ac{UE} $u_{2}$. %

Regarding the \ac{SNR} and \ac{SINR} perceived by the \acp{UE} in the \ac{DL}, their expressions are determined as follows. %

Consider a \ac{PRB} as the smallest allocable frequency unit, which consists of a number of adjacent subcarriers in the frequency domain. %
The system bandwidth is split into $K$ \acp{PRB}. %
Moreover, consider that \ac{UE} $u_{1}$ always connects directly to \ac{gNB} $b_{1}$, while \ac{UE} $u_{2}$ connects to \ac{gNB} $b_{2}$ either directly or through the \ac{NCR}. %

The power received by a node $r$ from a signal that was transmitted by a node $t$ at \ac{PRB} $k$, where the pair $(t,r)$ can be, for example, $(b_{1}, u_{1})$, $(b_{2}, u_{2})$, $(b_{2}, \text{NCR})$, $(\text{NCR}, u_{2})$, etc., is expressed as
\begin{equation}
p^{\text{R}}_{t,r,k} = \frac{p^{\text{T}}_{t,k} g^{\text{T}}_{t}g^{\text{R}}_{r}}{l_{t,r}},
\end{equation}
where $p^{\text{T}}_{t,k}$ is the power transmitted by node $t$ at \ac{PRB} $k$, $g^{\text{T}}_{t}$ is the transmit panel gain of node $t$, $g^{\text{R}}_{r}$ is the receive panel gain of node $r$, and $l_{t,r}$ is the pathloss between nodes $t$ and $r$. %
Notice that, for the pairs $(t,r) \in \{(b_{1},u_{2}), (b_{2},u_{1}), (\text{NCR},u_{1})\}$, $p^{\text{R}}_{t,r,k}$ represents an interfering power, while,
for the pairs $(t,r) \in \{(b_{1},u_{1}), (b_{2},u_{2}), (b_{2},\text{NCR}), (\text{NCR},u_{2})\}$, $p^{\text{R}}_{t,r,k}$ represents the useful power. %

When \ac{gNB} $b_{2}$ serves \ac{UE} $u_{2}$ via \ac{NCR}, the \ac{NCR} transmit power on \ac{PRB} $k$ $p^{\text{T}}_{\text{NCR},k}$ is equal to the \ac{NCR} total receive power, i.e., the sum of the useful power $p^{\text{R}}_{b_{2},\text{NCR},k}$, the noise $p_{n}$ and the interfering power $p^{\text{I}}_{\text{NCR},k} = p^{\text{R}}_{b_{1},\text{NCR},k}$, amplified by a gain $g^{\text{NCR}}$, limited by the \ac{NCR} maximum transmit power per \ac{PRB} $p^{\text{NCR}}_{\text{MAX}}$. %
It can be expressed as
\begin{equation}
p^{\text{T}}_{\text{NCR},k} = \text{min} \left\{p^{\text{NCR}}_{\text{MAX}}, g^{\text{NCR}}\left(p^{\text{R}}_{b_{2},\text{NCR},k} + p_{n} +p^{\text{I}}_{\text{NCR},k}\right)\right\}.
\end{equation}

We consider that the \ac{NCR} gain can be either fixed or dynamic\footnote{Only \ac{NCR} fixed gain has been specified by the \ac{3GPP}. So, while we try to mimic the model specified by 3GPP Rel-18, the considered setup may have differences with Rel-18 NCR.}. %
When using the dynamic gain, the \ac{NCR} transmit power $p^{\text{T}}_{\text{NCR},k}$ is always equal to the \ac{NCR} maximum transmit power $p^{\text{NCR}}_{\text{MAX}}$. %
For this, the dynamic gain is defined as the ratio between the \ac{NCR} maximum transmit power, i.e., $p^{\text{NCR}}_{\text{MAX}}$, and \ac{NCR} total receive power, i.e., the sum of the useful power $p^{\text{R}}_{b_{2},\text{NCR},k}$, the noise $p_{n}$ and the interfering power $p^{\text{I}}_{\text{NCR},k} = p^{\text{R}}_{b_{1},\text{NCR},k}$. %
We remark that the \ac{NCR} does not need to know the values of $p^{\text{R}}_{b_{2},\text{NCR},k}$, $p^{\text{I}}_{\text{NCR},k}$ and $p_{n}$ separately. %
It just need to know their sum, which is known since it is the total input power that it receives. %
Thus, the dynamic gain can be expressed as
\begin{equation}
g^{\text{NCR}}_{\text{DYN}} = \frac{p^{\text{NCR}}_{\text{MAX}}}{p^{\text{R}}_{b_{2},\text{NCR},k} + p^{\text{I}}_{\text{NCR},k} + p_{n}}.
\end{equation}

Regarding the fixed gain, i.e., $g^{\text{NCR}}_{\text{FIX}}$, it amplifies the total received power with a fixed value, e.g., $90$~dB. %
When the fixed gain implies an \ac{NCR} total transmit power higher than its ceiling $p^{\text{NCR}}_{\text{MAX}}$, the \ac{NCR} works in a saturated mode. %
In this mode, it behaves similar to the dynamic gain, where the \ac{NCR} transmit power $p^{\text{T}}_{\text{NCR},k}$ is equal to the \ac{NCR} maximum transmit power $p^{\text{NCR}}_{\text{MAX}}$. 

Thus, the \ac{SNR} and \ac{SINR} perceived by \ac{UE} $u_{i}$ at \ac{PRB} $k$ are, respectively, given by
\begin{equation}
\rho_{u_{i}, k} = \frac{p^{\text{R}}_{u_{i}, k}}{p^{\text{N}}_{u_{i}}},
\end{equation}
and
\begin{equation}
\eta_{u_{i}, k} = \frac{p^{\text{R}}_{u_{i}, k}}{p^{\text{I}}_{u_{i},k} + p^{\text{N}}_{u_{i}}},
\end{equation}
where $p^{\text{N}}_{u_{i}}$ is the receiver noise, $p^{\text{R}}_{u_{i}, k}$ is the useful power received by $u_{i}$ at \ac{PRB} $k$ and $p^{\text{I}}_{u_{i},k}$ is the interference suffered by $u_{i}$ at \ac{PRB} $k$. %

Regarding $p^{\text{R}}_{u_{i}, k}$, on the one hand, for $u_{1}$, the only useful signal received is the one coming from $b_{1}$, thus
\begin{equation}
p^{\text{R}}_{u_{1}, k} = p^{\text{R}}_{b_{1}, u_{1},k}. %
\end{equation}

On the other hand, $u_{2}$ may receive two components of useful signal, one coming directly from $b_{2}$ and other being amplified and forwarded by the \ac{NCR}. %
Thus, 
\begin{eqnarray}
p^{\text{R}}_{u_{2}, k} &=& p^{\text{R}}_{b_{2}, u_{2},k} + p^{\text{R}}_{\text{NCR}, u_{2},k} \nonumber \\
&=& \frac{p^{\text{T}}_{b_{2},k} g^{\text{T}}_{b_{2}}g^{\text{R}}_{u_{2}}}{l_{b_{2},u_{2}}} + \frac{\left(   p^{\text{R}}_{b_{2},\text{NCR},k}g^{\text{NCR}}\right) g^{\text{T}}_{\text{NCR}}g^{\text{R}}_{u_{2}}}{l_{\text{NCR},u_{2}}} \nonumber \\
&=& \frac{p^{\text{T}}_{b_{2},k} g^{\text{T}}_{b_{2}}g^{\text{R}}_{u_{2}}}{l_{b_{2},u_{2}}} + \frac{\left(\frac{p^{\text{T}}_{b_{2},k} g^{\text{T}}_{b_{2}}g^{\text{R}}_{\text{NCR}}}{l_{b_{2},\text{NCR}}}\right)g^{\text{NCR}} g^{\text{T}}_{\text{NCR}}g^{\text{R}}_{u_{2}}}{l_{\text{NCR},u_{2}}} \nonumber \\
&=& \frac{p^{\text{T}}_{b_{2},k} g^{\text{T}}_{b_{2}}g^{\text{R}}_{u_{2}}}{l_{b_{2},u_{2}}} + \frac{p^{\text{T}}_{b_{2},k} g^{\text{T}}_{b_{2}}g^{\text{R}}_{\text{NCR}}g^{\text{NCR}} g^{\text{T}}_{\text{NCR}}g^{\text{R}}_{u_{2}}}{l_{b_{2},\text{NCR}}l_{\text{NCR},u_{2}}}. \label{eq:useful_power}
\end{eqnarray} 

Concerning $p^{\text{I}}_{u_{i},k}$, with \ac{NCR}, there are two sources of interference for $u_{i}$, which are the signal from $b_{j}$, with $i \neq j$, and this same signal amplified by the \ac{NCR}, thus %
\begin{eqnarray}
p^{\text{I}}_{u_{i},k} &=& \frac{p^{\text{T}}_{b_{j},k} g^{\text{T}}_{b_{j}}g^{\text{R}}_{u_{i}}}{l_{b_{j},u_{i}}} + \frac{p^{\text{T}}_{b_{j},k} g^{\text{T}}_{b_{j}}g^{\text{R}}_{\text{NCR}}g^{\text{NCR}} g^{\text{T}}_{\text{NCR}}g^{\text{R}}_{u_{i}}}{l_{b_{j},\text{NCR}}l_{\text{NCR},u_{i}}}.
\end{eqnarray}

Finally, similarly to what has already been presented, regarding $p^{\text{N}}_{u_{i}}$, we have
\begin{eqnarray}
p^{\text{N}}_{u_{1}} &=& p_{n} \text{ and }\\
p^{\text{N}}_{u_{2}} &=& p_{n}\left(1 + \frac{g^{\text{NCR}} g^{\text{T}}_{\text{NCR}}g^{\text{R}}_{u_{2}}}{l_{\text{NCR},u_{2}}}\right). \label{eq:amplified_noise}
\end{eqnarray}

\section{Performance Evaluation} \label{CHP:NCR_Interf_SEC:Perf_Eval}

\subsection{Simulation Assumptions} \label{CHP:NCR_Interf_SEC:Sim_Assumptions}

The simulations were conducted at 28 GHz. 
The frequency domain was split into \acp{PRB} consisting of 12 consecutive subcarriers, with subcarrier spacing of 60 kHz. %
It was adopted the \ac{RR} scheduler for allocating the \acp{PRB}. 

Concerning the time domain, it was split into slots composed of 14~\ac{OFDM} symbols. %
Each slot had a duration of 0.25 ms. %
A \ac{TDD} scheme was adopted, where downlink and uplink slots were alternated in time. %

Regarding the channel model, the adopted one is based on the \ac{3GPP} channel model standardized in~\cite{3gpp.38.901} and its implementation is described in~\cite{Pessoa2019}. %
In this channel model, it is considered a distance-dependent pathloss, a lognormal shadowing component, a small-scale fading, and it is spatially and time consistent. The link types are described in \TabRef{TABLE:NCR_Interf_SEC:LINKS}. %
The \acp{gNB} and the \ac{NCR} transmissions were performed with a \ac{DFT} codebook based beamforming, where for each transmission a beam management was performed in order to identify the best transmitter beam to be used when serving the \acp{UE}. %

\begin{table}[tb]
	\caption{Characteristic of the links}
	\begin{center}
		{
			\begin{tabular}{ccc}\hline
				\textbf{Link}  & \textbf{Scenario}  & \textbf{LOS/NLOS} \\\hline
				gNB - UE       & Urban Macro        & NLOS \\
				gNB - NCR      & Urban Macro        & NLOS \\
				NCR - UE       & Urban Micro        & LOS \\\hline
			\end{tabular}
		}
	\end{center}
	\label{TABLE:NCR_Interf_SEC:LINKS}
\end{table}


It was used a \ac{CQI}/\ac{MCS} mapping curve standardized in~\cite{3gpp.38.214} with a target \ac{BLER} of 10~\%.  %
An outer loop strategy was considered to avoid the increase of the \ac{BLER}, i.e., when a transmission error occurred, the estimated \ac{SINR} decreased 1 dB, however, when a transmission occurred without error, the estimated SINR had its value added by 0.1~dB. %
The most relevant simulation parameters are summarized in Tables~\ref{TABLE:NCR_Interf_SEC:Entities-characteristics} and \ref{TABLE:NCR_Interf_SEC:Simul_Param}. %

\begin{table*}[!t]
	\centering
	\caption{Entities characteristics.}
	\label{TABLE:NCR_Interf_SEC:Entities-characteristics}
	\begin{tabularx}{\textwidth}{lXXX}
		\hline
		\textbf{Parameter} & \textbf{Macro \ac{gNB}} & \textbf{\ac{NCR}} & \textbf{\ac{UE}} \\
		\hline
		Height & 25 m & 10 m & 1.5 m \\
		Transmit power & 16.8 dBm & 13.8 dBm & 5.8 dBm \\
		Antenna array & URA $8\times 8$ & URA $8\times 8$ (2 panels) & Single Antenna \\
		Antenna element pattern & \ac{3GPP} 3D~\cite{3gpp.38.901} & \ac{3GPP} 3D~\cite{3gpp.38.901} & Omni \\
		Max. antenna element gain & 8 dBi & 8 dBi & 0 dBi \\
		\hline
	\end{tabularx}
\end{table*}

\begin{table}
	\centering
	\setlength{\tabcolsep}{1ex}
	\caption{Simulation parameters.}
	\label{TABLE:NCR_Interf_SEC:Simul_Param}
	\begin{tabularx}{0.99\columnwidth}{>{\raggedright\arraybackslash}X>{\raggedright\arraybackslash}X}
		\hline
		\textbf{Parameter} & \textbf{Value} \\
		\hline
		Carrier frequency & 28 GHz\\
		Subcarrier spacing & 60 kHz\\
		Number of subcarriers per \acs{RB} &  $12$\\
		Number of \acsp{RB} & $1$\\
		Slot duration & 0.25 ms \\
		OFDM symbols per slot & $14$ \\
		Channel generation procedure & As described in~\cite[Fig.7.6.4-1]{3gpp.38.901}\\
		Path loss  & Eqs. in~\cite[Table 7.4.1-1]{3gpp.38.901}\\
		Fast fading & As described in~\cite[Sec.7.5]{3gpp.38.901} and \cite[Table 7.5-6]{3gpp.38.901} \\
		AWGN density power per subcarrier & -174 dBm/Hz\\
		Noise figure &  9 dB\\
		\acs{CBR} packet size & $3072$ bits \\
		\Acl{ISD} & 400 m \\
		Distance between \ac{gNB} and \ac{UE} & 150 m \\
		\hline
	\end{tabularx}
\end{table}

\subsection{Simulation Results} \label{CHP:NCR_Interf_SEC:Sim_Res}

\begin{figure}[t]
	\centering
	\includegraphics[scale=0.38]{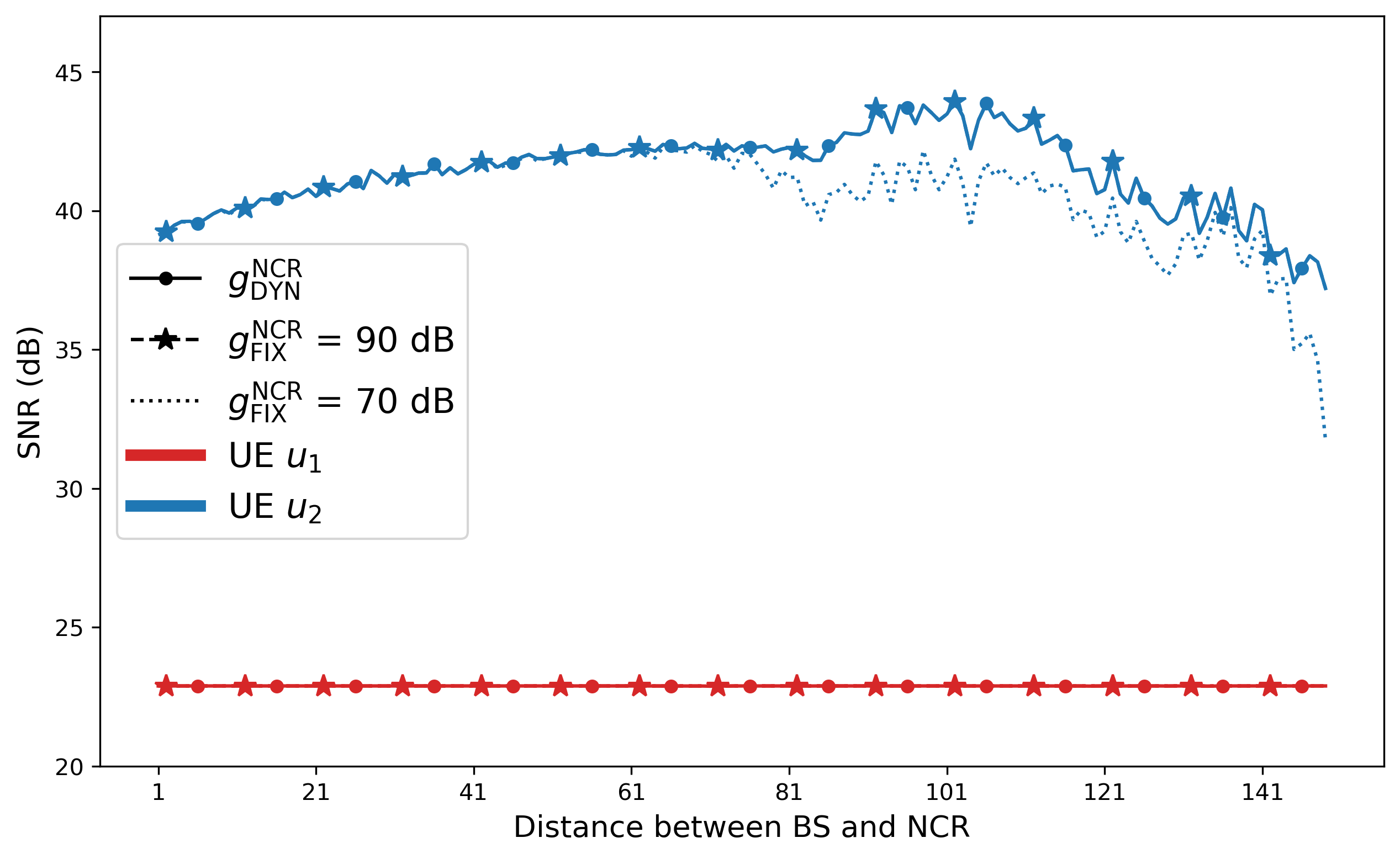}
	\caption{Impact of \acs{NCR} position on the \ac{SNR} (quantile 90\%) of  both \acp{UE} for two types of \ac{NCR} gain, i.e., dynamic and fixed.}
	\label{FIG:NCR_Interf_SEC:SNR_quantile9}
\end{figure}

\begin{figure}[t]
	\centering
	\includegraphics[scale=0.38]{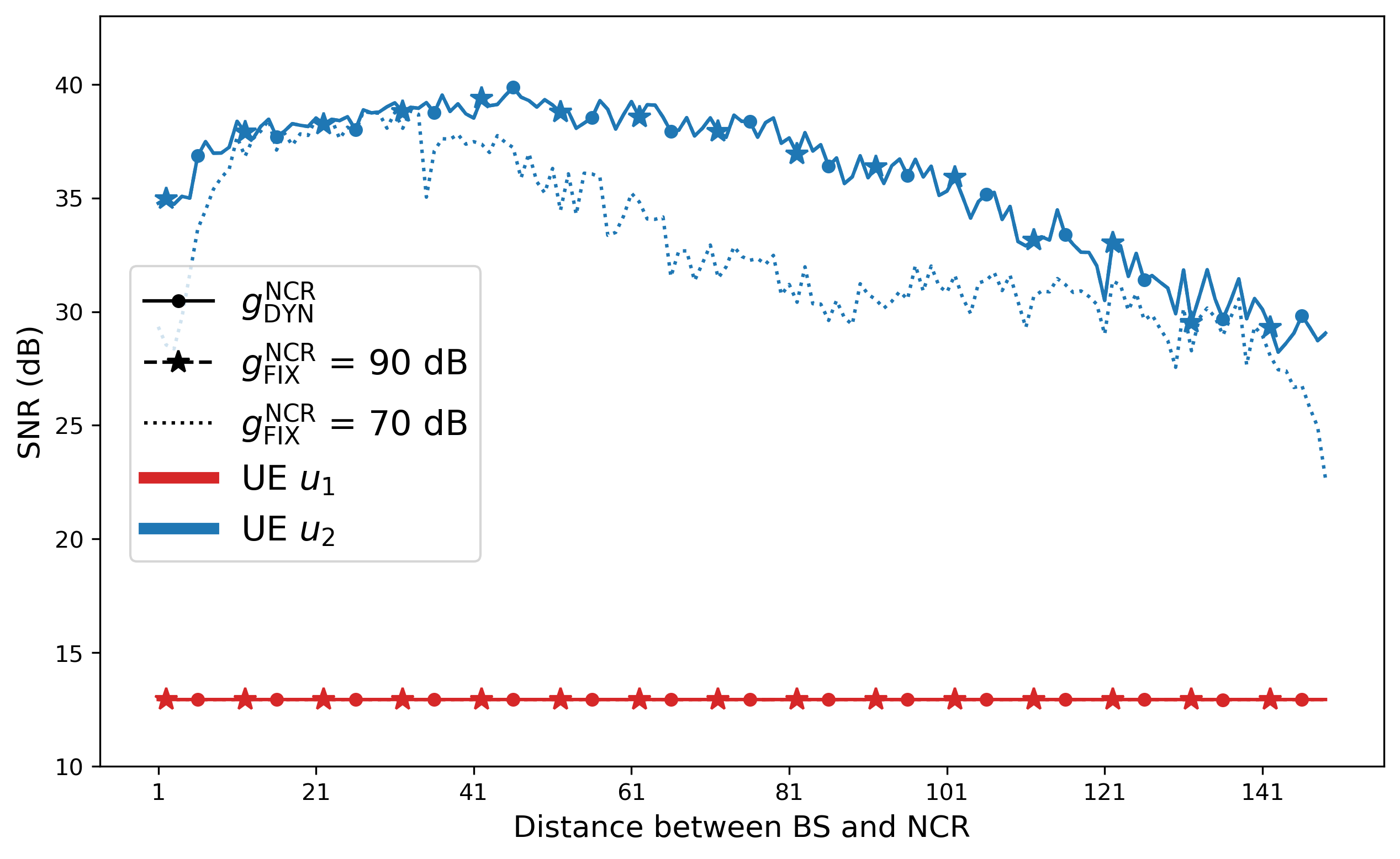}
	\caption{Impact of \acs{NCR} position on the \ac{SNR} (quantile 10\%) of  both \acp{UE} for two types of \ac{NCR} gain, i.e., dynamic and fixed.}
	\label{FIG:NCR_Interf_SEC:SNR_quantile1}
\end{figure}

Figures \ref{FIG:NCR_Interf_SEC:SNR_quantile9} and \ref{FIG:NCR_Interf_SEC:SNR_quantile1} show the impact of \ac{NCR} position on the \ac{SNR} (quantiles 0.9 and 0.1, respectively) of  \acp{UE} $u_1$ and $u_2$. %
Both figures present the results for the two possibilities of the \ac{NCR} gain: dynamic and fixed. %
For the fixed case, we considered two gain values: 70~dB and 90~dB. %
Considering the positions of $u_{2}$ and $b_{2}$ fixed, the x-axis represents the possible distance between the \ac{NCR} and $b_{2}$. %
As expected, the \ac{SNR} of $u_1$ does not depend of the position of the \ac{NCR}. %
Also, notice that the \ac{SNR} of $u_2$, in the beginning, increases and, after a certain distance, starts to decrease. %
This is due to the product of the two pathlosses in (\ref{eq:useful_power}), i.e., $l_{b_{2},\text{NCR}}l_{\text{NCR},u_{2}}$. %
Thus, unlike what one could expect, deploying a \ac{NCR} closer to the serving \ac{UE} does not always mean a better connection. %
In other words, one could expect that the closer a \ac{UE} is to a \ac{NCR} the higher its \ac{SNR} would be due to the shorter distance, and so lower path loss, however as shown in Figs.~\ref{FIG:NCR_Interf_SEC:SNR_quantile9} and \ref{FIG:NCR_Interf_SEC:SNR_quantile1}, this is not true. %

Due to the symmetry of the scenario presented in \FigRef{FIG:NCR_Interf_SEC:Scenario}, without the \ac{NCR}, $u_1$ and $u_2$ should present similar values of \ac{SNR} and \ac{SINR}. %
Thus, by comparing the curves in \FigRef{FIG:NCR_Interf_SEC:SNR_quantile9} and \FigRef{FIG:NCR_Interf_SEC:SNR_quantile1} related to $u_1$ and $u_2$, we can see that the deployment of an \ac{NCR} considerably improves the \ac{SNR} perceived by $u_2$. %
More specifically, the \ac{SNR} perceived by $u_2$ increased in at least 15~dB due to the deployment of the \ac{NCR}. %
Moreover, notice that the case with fixed \ac{NCR} gain equal to 90~dB presented results similar to the dynamic case, which means that for 90~dB, the \ac{NCR} operated in its saturated mode. %

\begin{figure}[t]
	\centering
	\includegraphics[scale=0.38]{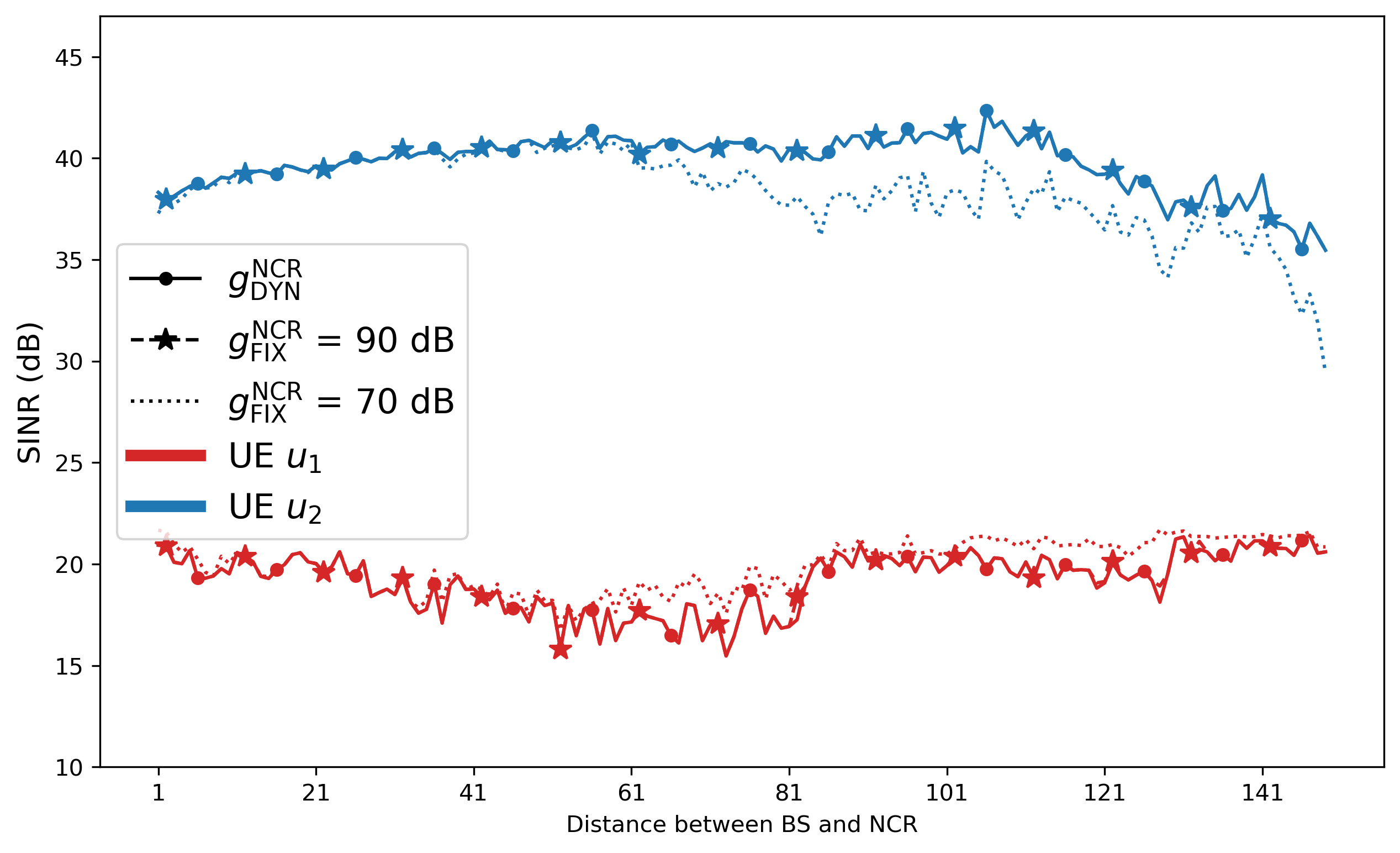}
	\caption{Impact of \acs{NCR} position on the \ac{SINR} (quantile 90\%) of both \acp{UE} for two types of \ac{NCR} gain, i.e., dynamic and fixed.}
	\label{FIG:NCR_Interf_SEC:SINR_quantile9}
\end{figure}

\begin{figure}[t]
	\centering
	\includegraphics[scale=0.38]{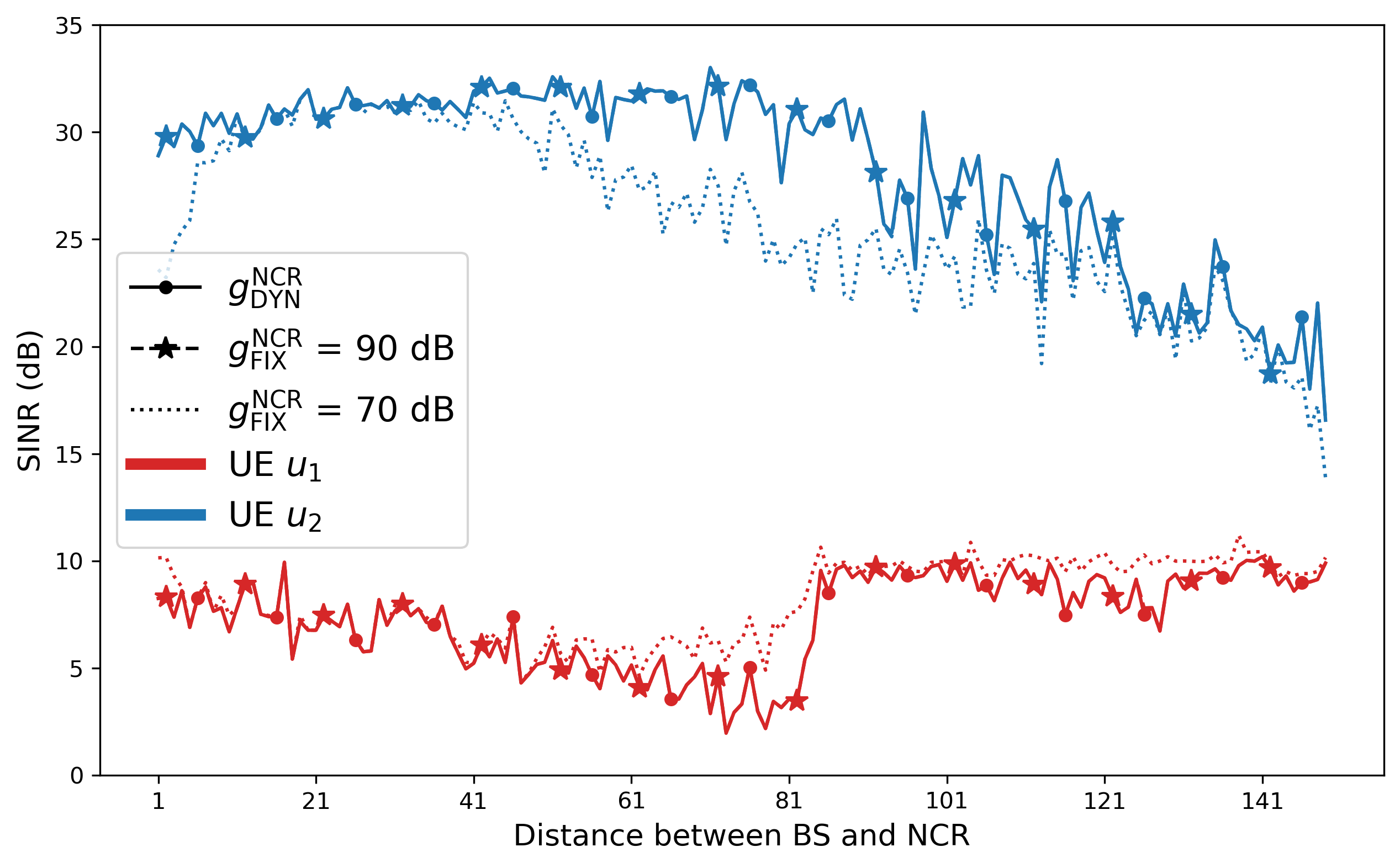}
	\caption{Impact of \acs{NCR} position on the \ac{SINR} (quantile 10\%) of both \acp{UE} for two types of \ac{NCR} gain, i.e., dynamic and fixed.}
	\label{FIG:NCR_Interf_SEC:SINR_quantile1}
\end{figure}

Figures~\ref{FIG:NCR_Interf_SEC:SINR_quantile9} and \ref{FIG:NCR_Interf_SEC:SINR_quantile1} are similar to Figs.~\ref{FIG:NCR_Interf_SEC:SNR_quantile9} and \ref{FIG:NCR_Interf_SEC:SNR_quantile1}, the main difference is that Figs.~\ref{FIG:NCR_Interf_SEC:SINR_quantile9} and \ref{FIG:NCR_Interf_SEC:SINR_quantile1} focus on  \ac{SINR} instead of \ac{SNR}. %

\begin{figure}[t]
	\centering
	\includegraphics[scale=0.39]{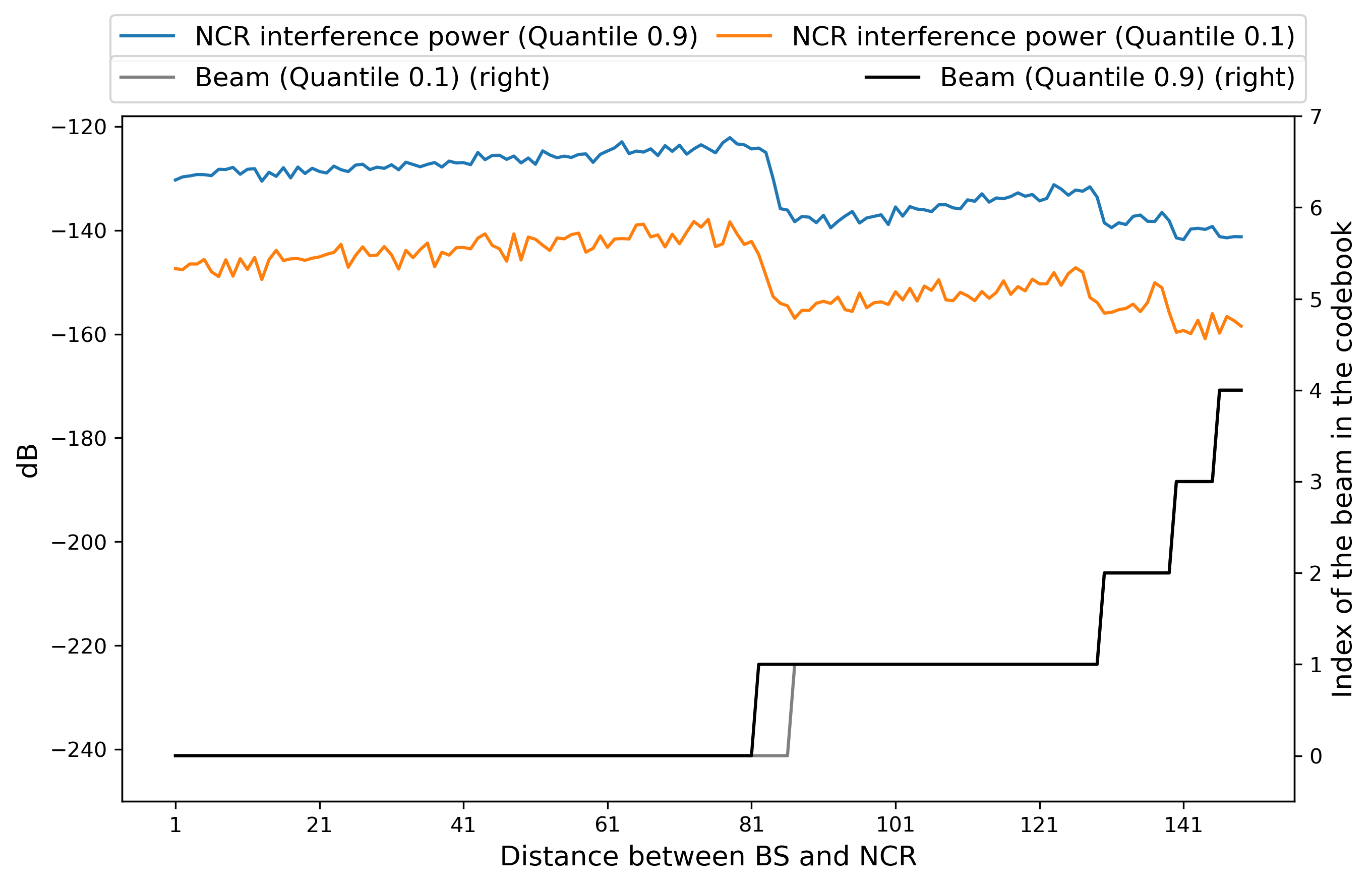}
	\caption{Comparison between the interference and amplified noise by \ac{NCR} in the $u_1$ and the change in the codebook indexes of the \ac{NCR}-\ac{Fwd}.}
	\label{FIG:NCR_Interf_SEC:BEAM_NCR}
\end{figure}

Notice, in Figs.~\ref{FIG:NCR_Interf_SEC:SINR_quantile9} and \ref{FIG:NCR_Interf_SEC:SINR_quantile1}, the trend discontinuity on the \ac{SINR} of $u_1$ when the distance between the \ac{NCR} and $b_{2}$ is approximately 81~m. %
This is explained by the change in the interference coming from the \ac{NCR}. %
More specifically, around that distance the beam used by the \ac{NCR} to serve $u_{2}$ changes creating a new interference pattern on $u_{1}$. %
\FigRef{FIG:NCR_Interf_SEC:BEAM_NCR} illustrates this behavior. %

\FigRef{FIG:NCR_Interf_SEC:BEAM_NCR} presents the impact of the distance between $b_{2}$ and the \ac{NCR} on the interference suffered by $u_{1}$ (y-axis of left-hand side) and on the beam index that is used to serve $u_{2}$ (y-axis of right-hand side). %
In this figure, we can notice that the interference suffered by $u_{1}$ has a discontinuity when the distance between $b_2$ and the \ac{NCR} is around 81~m. %
Around this position, the \ac{NCR} beam serving $u_{2}$ changes from 0 to 1. %
Beam~0 points in a direction closer to $u_1$ than Beam~1, that is why when changing from Beam~0 to Beam~1, the interference decreases. %
This can be seen as a spatial filtering. %

Furthermore, notice, in \FigRef{FIG:NCR_Interf_SEC:BEAM_NCR}, that within the distance ranges 0~m to 81~m and 81~m to 150~m, the interference suffered by $u_1$ increases when the \ac{NCR} distance between $b_{2}$ and the \ac{NCR} increases. %
This is explained by the approximation of the \ac{NCR} to $u_{1}$. %



\begin{figure}[t]
	\centering
	\includegraphics[scale=0.39]{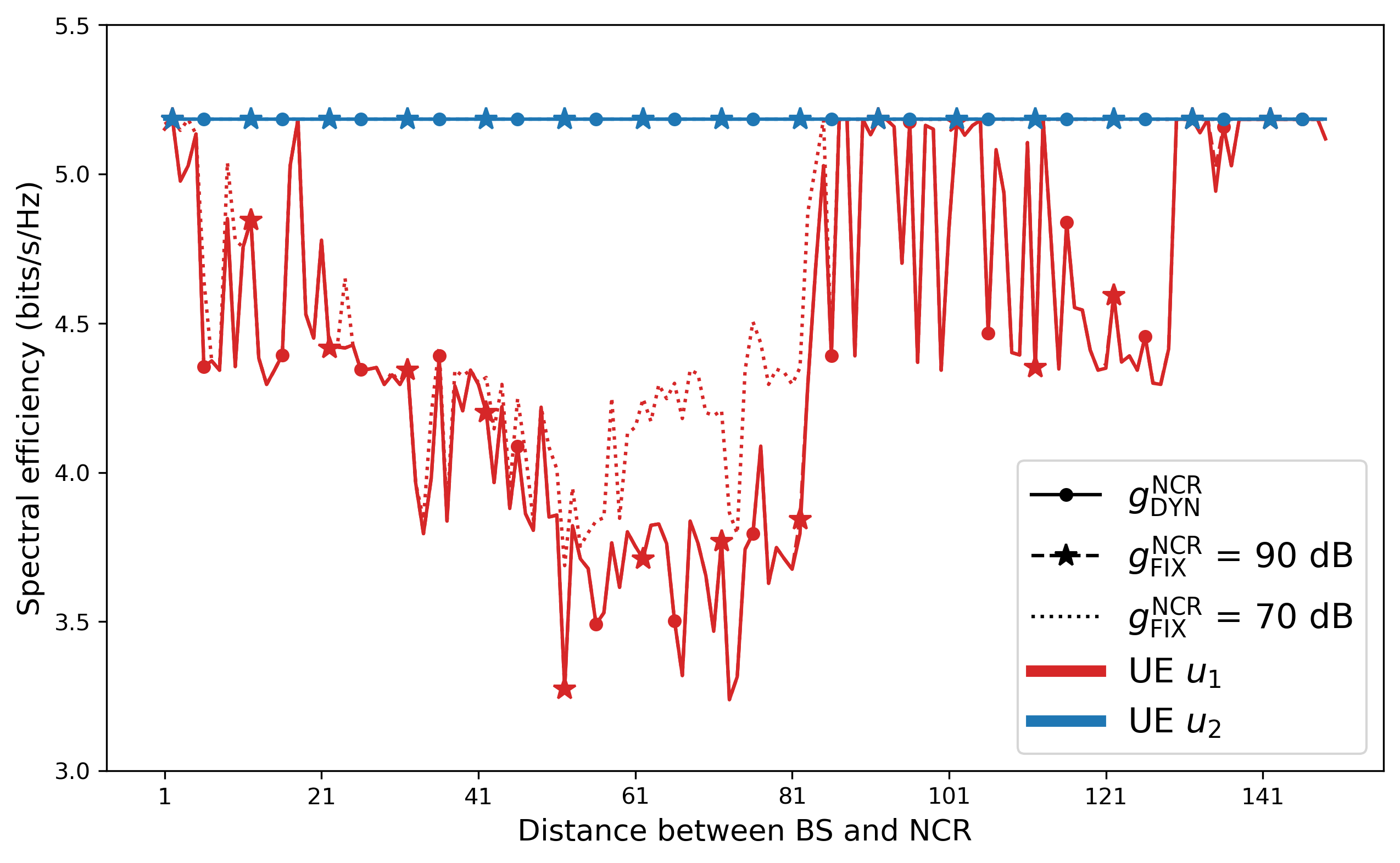}
	\caption{Spectral efficiency (quantile 90\%) of the both \acp{UE} and with different types of \ac{NCR} gain.}
	\label{FIG:NCR_Interf_SEC:Spectral_quantile9}
\end{figure}

\begin{figure}[t]
	\centering
	\includegraphics[scale=0.39]{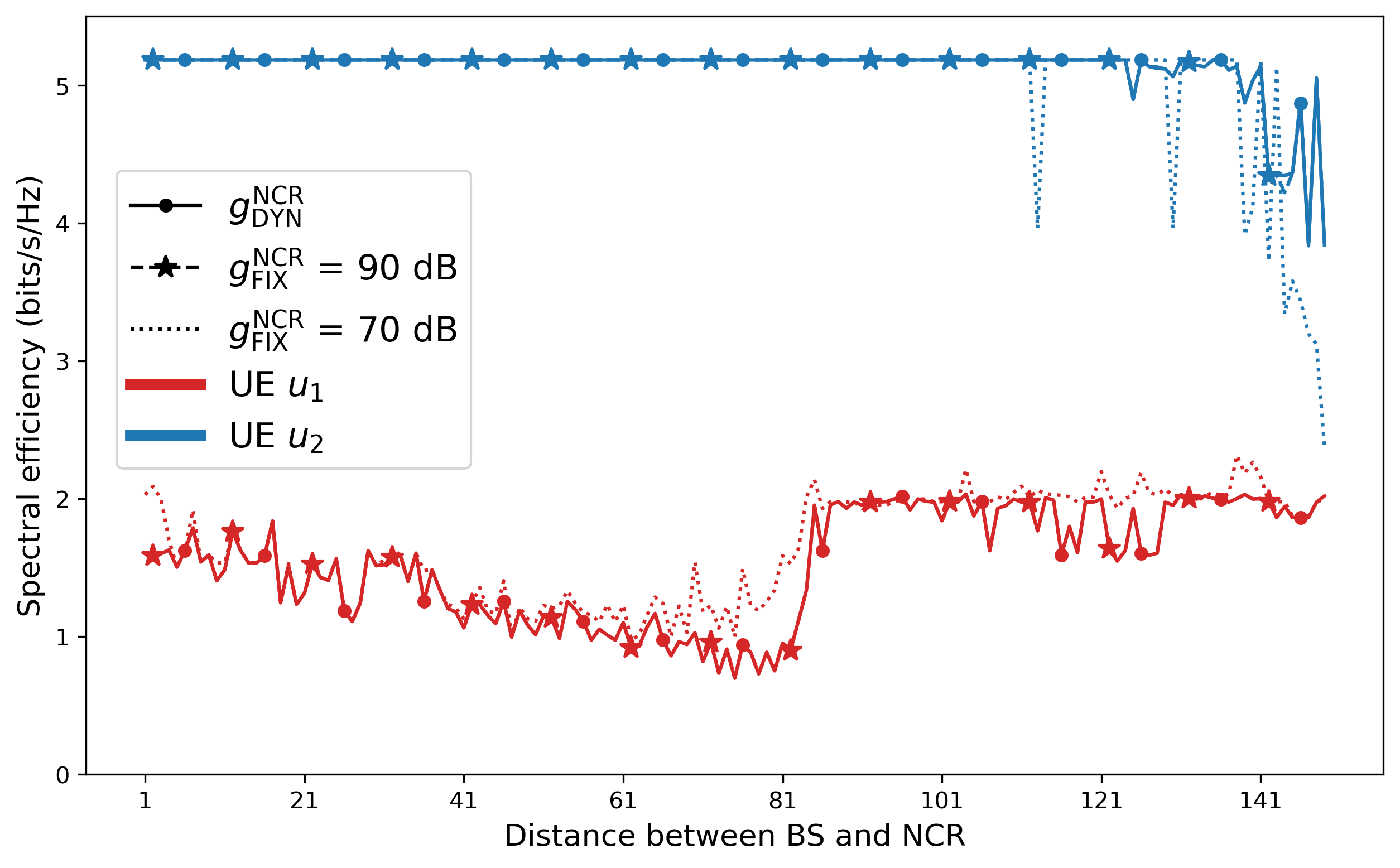}
	\caption{Spectral efficiency (quantile 10\%) of the both \acp{UE} and with different types of \ac{NCR} gain.}
	\label{FIG:NCR_Interf_SEC:Spectral_quantile1}
\end{figure}

Finally, Figs.~\ref{FIG:NCR_Interf_SEC:Spectral_quantile9} and \ref{FIG:NCR_Interf_SEC:Spectral_quantile1} present the impact of the distance between $b_2$ and the \ac{NCR} on the spectral efficiency of the transmissions to $u_{1}$ and $u_{2}$. %
The maximum spectral efficiency that a transmission could achieve is:
\begin{eqnarray}
&&\frac{\text{subcarriers per PRB}\cdot \text{symbols per PRB} \cdot \text{max. code rate}}{\text{PRB time} \cdot \text{PRB bandwidth}} \nonumber \\
&=& \frac{12 \cdot 14 \cdot 5.5547}{0.25\cdot 10^{-3} \cdot 12 \cdot 60\cdot 10^{3}} = 5.18 \text{ bits/s/Hz},
\end{eqnarray}
where 5.5547 corresponds to the code rate associated to the \ac{CQI} index 15 in~\cite{3gpp.38.214}. %
Remark that $u_2$ achieves the maximum spectral efficiency in the majority of the considered cases, while $u_{1}$ achieves lower values, but that are still high enough to allow the communication. %


\section{Conclusions} \label{CHP:NCR_Interf_SEC:Conclusion}

The paper presented a system level evaluation analyzing the performance improvement due to the deployment of a \ac{NCR} on a given cell and its interference impact on neighbor cells. %
As expected, we have seen that the \ac{NCR} improves the link quality of its serving \ac{UE}. %
However, unlike what one could expect, deploying a \ac{NCR} closer to its serving \ac{UE} does not necessarily mean a better connection. %
There is a trade-off given by the product between its  distance to its serving \ac{gNB} and its distance to the \ac{UE} that it is serving. %
We have also seen that the interference caused on neighbor cells can be mitigated by spatial filtering by means of appropriate beam management. %


\printbibliography
\end{document}